\documentclass[]{aastex631}

\usepackage{graphicx,amssymb,amsmath,times,xcolor}
\usepackage{longtable}
\usepackage{bm}
\usepackage{url}
\usepackage[varg]{txfonts}
\usepackage[T1]{fontenc}
\usepackage{natbib}
\bibpunct{(}{)}{;}{a}{}{,} 
\usepackage[normalem]{ulem}

\definecolor{operamauve}{rgb}{0.718, 0.518, 0.655}

\definecolor{red}{rgb}{1., 0., 0.}

\definecolor{blue}{rgb}{0.2, 0.2, 1.}

\graphicspath{{./}{figures/}}

\begin{document}
\title{Light curve's recovery with Rubin-LSST:\\
II. UnVEiling the darknesS of The gAlactic buLgE (VESTALE) with RR Lyrae }
\author[0000-0003-4132-1209]{Di Criscienzo,M. }
\affiliation{INAF – Osservatorio Astronomico di Roma, Via Frascati 33, I-00078 Monte Porzio Catone, Roma, Italy}
\author{Leccia, S.}
\affiliation{INAF-Osservatorio Astronomico di Capodimonte, Salita Moiariello 16, 80131 Naples, Italy}
\author{Braga, V.}
\affiliation{INAF – Osservatorio Astronomico di Roma, Via Frascati 33, I-00078 Monte Porzio Catone, Roma, Italy}
\author{Musella, I.}
\affiliation{INAF-Osservatorio Astronomico di Capodimonte, Salita Moiariello 16, 80131 Naples, Italy}
\author{Bono, G.}
\affiliation{Department of Physics, Universit\'a di Roma Tor Vergata, Via della Ricerca Scientifica 1, 00133 Roma, Italy}
\author{Dall'Ora}
\affiliation{INAF-Osservatorio Astronomico di Capodimonte, Salita Moiariello 16, 80131 Naples, Italy}
\author{Fiorentino, G.}
\affiliation{INAF – Osservatorio Astronomico di Roma, Via Frascati 33, I-00078 Monte Porzio Catone, Roma, Italy}
\author{Marconi, M.}
\affiliation{INAF-Osservatorio Astronomico di Capodimonte, Salita Moiariello 16, 80131 Naples, Italy}
\author{Molinaro, R.}
\affiliation{INAF-Osservatorio Astronomico di Capodimonte, Salita Moiariello 16, 80131 Naples, Italy}
\author{Ripepi, V.}
\affiliation{INAF-Osservatorio Astronomico di Capodimonte, Salita Moiariello 16, 80131 Naples, Italy}
\author{Girardi, L}
\affiliation{Osservatorio Astronomico di Padova – INAF, Vicolo dell’Osservatorio 5, I-35122 Padova, Italy}
\author{Mazzi, A.}
\affiliation{Universita di Bologna}
\author{Pastorelli, G.}
\affiliation{Dipartimento di Fisica e Astronomia Galileo Galilei, Universit`a di Padova, Vicolo dell’Osservatorio 3, I-35122 Padova, Italy}
\author{Trabucchi, M.}
\affiliation{Dipartimento di Fisica e Astronomia Galileo Galilei, Universit`a di Padova, Vicolo dell’Osservatorio 3, I-35122 Padova, Italy}
\author{N. Matsunaga}
\affiliation{Department of Astronomy, School of Science, The University of Tokyo, 7-3-1 Hongo, Bunkyo-ku, Tokyo 113-0033, Japan}
\author{Monelli, M.}
\affiliation{Instituto de Astrofísica de Canarias, Calle Via Lactea s/n, 38205 La Laguna, Tenerife, Spain}
\author{Saha, A.}
\affiliation{National Optical Astronomy Observatory, 950 N. Cherry Avenue, Tucson, AZ 85719, USA}
\author{Vivas, K. A.}
\affiliation{Cerro Tololo Inter-American Observatory/NSF’s NOIRLab, Casilla 603, La Serena, Chile}
\author{Zanmar Sanchez, R.}
\affiliation{INAF-Osservatorio Astronomico di Capodimonte, Salita Moiariello 16, 80131 Naples, Italy}

\begin{abstract}
This work is part of VESTALE, a project initiated within the LSST Cadence Strategy Optimization Process \citep{vestale}. Its goal is to explore the potential of Rubin-LSST observations aimed at the Galaxy's bulge (Bulge) for studying RR Lyrae  stars (RRL).
Observation and analysis of RR Lyrae stars in the Bulge are crucial for tracing the old population of the central part of our galaxy and reconstructing the history of Bulge formation. Based on observations conducted with CTIO/DECam by \citet{saha19} towards the Baade Window, our simulations demonstrate that early Rubin-LSST observations will enable the recovery of RR Lyrae light curves at Galactic center distances with sufficient precision. This will allow us to utilize theoretical relations from \citet{marconi2022} to determine their distances and/or metallicity, following the REDIME algorithm introduced in \citet{bono}.
We show how reddening and crowding affect our simulations and highlight the importance of considering these effects when deriving pulsation parameters (luminosity amplitudes, mean magnitudes) based on the light curves especially if the goal is to explore the opposite side of the Bulge through the observation of its RRL.
The simulations discussed in this investigation were conducted to support the SCOC's decision to observe this important sky region since it has only recently been decided to include part of the Bulge as a target within the LSST main survey.
 
\end{abstract}

\section{introduction}
Although detailed photometric investigations of the Bulge date back to more than one century ago thanks to the pioneering works of  Shapley and Baade, its formation and evolution remain a debated topic \citep{kunder2022}.\\
Old, metal-poor stars in the inner Galaxy need to be adequately accounted for when discussing processes that gave rise to the formation of the Galactic bar/bulge.
In general, RRL can be easily recognized among these old stars due to the coupling between the shape of the light curve 
and the pulsation period and for this reason, they are often used as tracers of old stellar populations in different Galactic and extragalactic environments.\\
Even if the Bulge is a challenging environment due to the effect of high reddening and stellar crowding, several studies successfully investigated the RRL's light curves in the last decade. We can mention for example  MACHO, EROS and OGLE, optical surveys dedicated to microlensing, which discovered thousands of variable stars in the Bulge \citep{MACHO,EROS,s2019} or the VVV survey \citep{vvv2018} which, working in the near-infrared bands, has the advantage of being less sensitive to reddening than those in the optical passbands and indeed, they discovered one thousand  
new RRL in the Galactic Centre (GC).\\
Recent studies based on these large samples \citep[see][for a complete review]{kunder2022} show that the Bulge RRL have a metallicity distribution function centered around [Fe / H] $\sim$ -1.4 , but covering a very broad range in iron abundance \citep{walker,savino}. 
This means that RRL are solid tracers to investigate the early chemical enrichment of the Bulge \citep{debattista}.
As for the kinematic properties, there still needs to be a consensus as to whether the Bulge RRL spatially trace out the bar or if they are more consistent with a more spheroidal bulge population. In particular  different RRL photometric surveys in different wavelengths present differing results.
To further complicate the situation many of the results from current investigations are affected by several limitations: i) deep and homogeneous optical investigations have been limited to low reddening windows; ii) the Bulge, and in particular the inner part, are among the densest Galactic stellar fields. This means that both optical and NIR seeing limited observations are hampered by two limitations: i) even relatively bright sources are affected by blending; ii) the limiting magnitude is quite shallow, due to the confusion limit.\
This paper will focus on what the soon-to-start Rubin Legacy Survey of Space and Time (Rubin-LSST)  can do for the RRL observation in the Bulge. 
In particular  in  \citet{marconi2022} we have derived new theoretical Color-Color and Period–Luminosity–Metallicity relations for RRL in the Rubin-LSST filters based on a recently computed extensive set of nonlinear convective pulsation models for RRL stars, covering a broad range of metal content. As demonstrated in \citet{bono}, these relationships offer an important tool for deriving the metallicity, distance, and reddening of these stars from the recovery of their light curves. However, an accurate recovery of RRL light curves is necessary to precisely determine pulsational parameters and thus fully exploit the Marconi et al. relationships.
As a part of the VESTALE (unVEil the darknesS of The gAlactic buLgE) project \citep{vestale}, in this paper we will simulate Rubin-LSST  observations of RRL in the Bulge direction, to examine the level of accuracy we can expect in the recovery of light curves and  to understand how deep into the darkness we can go. This provides an exceptional opportunity to determine the density profile of old stellar populations in the direction of the Bulge.
We remember that this particular footprint was only recently included in the main survey as a result of the cadence strategy optimization process and in response to the pressure from the scientific community interested in this important  part of the Galaxy \citep{street2023}.\\
To understand how much the decisions made at the end of Phase 2 of the Cadence Strategy Optimization Process will impact the accuracy in the recovery of Bulge RRL, in this paper we will use both tools released by LSST collaboration and our own.\\
The structure of the paper is the following.
In Section 2, we will review the decisions made by the Survey Cadence Optimization Committee (SCOC) that impact the Bulge, as summarized in \texttt{\texttt{baseline\_v3.3\_10yrs.db}}.\\
In section 3, we show the Rubin-LSST simulations of RRL's light curves (LC) discovered and studied by \cite{saha19} with CTIO/DECam. We analyze LC recovery in function of time to understand how much better Rubin-LSST will be able to perform and when the first useful results for studies of Bulge formation will begin to be obtained.\\
Section 4 investigates the effect of crowding on the recovery of RRL light curves and  discuss the importance of including crowding-related errors as data products in each release of Rubin-LSST data.\\
Considering that, in addition to extending Saha's work to a much broader sky area towards the Bulge, the survey will probe much deeper than any other survey of its kind in Section 5, we displace Saha's templates further away  and repeat the LC recovery to understand the extent of LSST's capabilities.\\
The Conclusions close the paper.\\

\section{The inner part of the  Galactic Bulge in Rubin-LSST Opsim simulations}
The first released simulations produced by the Operations Simulator (OpSim) within the Metrics Analysis Framework (MAF) omitted the consideration of the Bulge/Inner disk (and the Magellanic Clouds) in the primary survey due to issues related to crowding. Instead, these specific sky regions were encompassed by a series of \textit{ad hoc} mini-surveys, each comprising approximately one-fourth the number of visits projected by the main survey.\\
Thanks to the pressure of the scientific community \citep[see e.g.][]{olsen2018}, the SCOC in its final document (Phase 2 document, PSTN-055\footnote{\UrlFont{https://pstn-055.lsst.io/}}) added the suggestion to include part of the Bulge and Plane in the main survey called  Wide, Fast and Deep (WFD) survey.
In particular, the recommended footprint implemented in the last version  of the Baseline (\texttt{\texttt{baseline\_v3.0\_10yrs.db}}) included a $\sim$ 10$\%$ area increase, extending the WFD into low-dust extinction regions for extragalactic science, and expanded WFD-level coverage into high-priority regions for Galactic science — the Galactic Bulge and the Magellanic Clouds. 
The proposed solution is to cover the Galactic plane in two parts: WFD-level regions near the Bulge and other regions of interest (that are still to be decided in a definitive manner) and a lower level of coverage in the rest of the regions. The median number of visits per pointing in these areas over the lifetime of the survey (in any filter) is $\sim$ 800 visits per pointing in WFD-level areas and $\sim$ 250 in the high dust extinction (low coverage) Galactic plane regions. Actually there's still a preference for a distribution of visits in the rizy bands (20\% in each of these bands), while $\sim$ 7\% and 10\% of the visits will be in the u and g bands respectively.\\
Nevertheless, a few outstanding points remain open for further exploration and refinement.
A new task force that formed immediately after the conclusion of Phase 2 of the Cadence Strategy Optimization process, is currently working to provide input on the final design of the coverage of areas of the Galactic Plane/Bulge.
In particular the 4th Survey Cadence Optimization Workshop (  \texttt{https://www.youtube.com/watch?v=Z97ZTagYd8I}) has just concluded, and one of the requests from the SCOC at the margin of the meeting was to provide feedback on the latest simulations, especially from teams that refer to Stars, Milky Way \& Local Volume (SMWLV)  science collaboration interested in reconstruct the shape and evolution of Galaxy and its components with  LSST data.
This work partially aims to address this request.\\
In addition, this work provides us with the opportunity to explore, for the first time, the new simulations constructed considering the triple silver coating of the telescope's mirrors. We recall that the sensitivity has notably improved in grizy photometric bands between baseline version 3.0 (described in phase2 document ) and 3.3 and decreased by approximately 0.4 mag the u-band sensitivity.

\section{LSST recovery of  RR Lyrae observed by Saha et al. 2019 in Baade Window }
To simulate LSST observations of Bulge's RRL we  use PulsatingStarRecovery (PSR) metric \citep{dicriscienzo2023} that we build within the Cadence Survey Optimization process to analyze the recovery of the LSST simulated light curves.\\
We start from the area centered on the well-known Baade's Window  ((l, b) = (1.02, -3.92) deg) which has the advantage of being close to the direction of the Galactic center while remaining relatively transparent to dust. 
The further advantage of this area is that ugriz observations at different epochs with DECam imager on the Blanco 4m telescope at CTIO are available. DECam not only has similar filters, but also has a pixel scale very similar to LSSTCam. 
\citet{saha19} observed a single DECam field (FoV=3 deg$^2$) in Baade’s Window, obtaining between 70-90 epochs per band.
According to the last Opsim version (\texttt{baseline\_v3.3\_10yrs.db}) Baade's Window region will be observed by the main survey which means that it will receive an average of 800 visits over the 10 years of the survey in all the photometric bands. \\
Using these data  \cite{saha19} confirmed and analyzed almost 500 RRL, almost all ab-type. 
For each of the confirmed RRL the authors  provided the best template taken from the library of light curve templates set up from \cite{seasar2010}. We use those curves  as starting points to derive the simulated LSST light curves .\\
From the Saha's sample we only removed stars that have, in at least one photometric band, an average magnitude brighter than the saturation magnitude of Rubin-LSST (see Science book, \texttt{https://www.lsst.org/scientists/scibook}). As a result, the number of RRL that we simulated decreased from 474 to 439 stars.\\
We begin analyzing the recovery of RRL as they would appear if the LSST cadence followed one of the latest baselines released by SCOC (namely \texttt{baseline\_v3.3\_10yrs.db}). 
\begin{figure}
\includegraphics[width=0.5\linewidth]{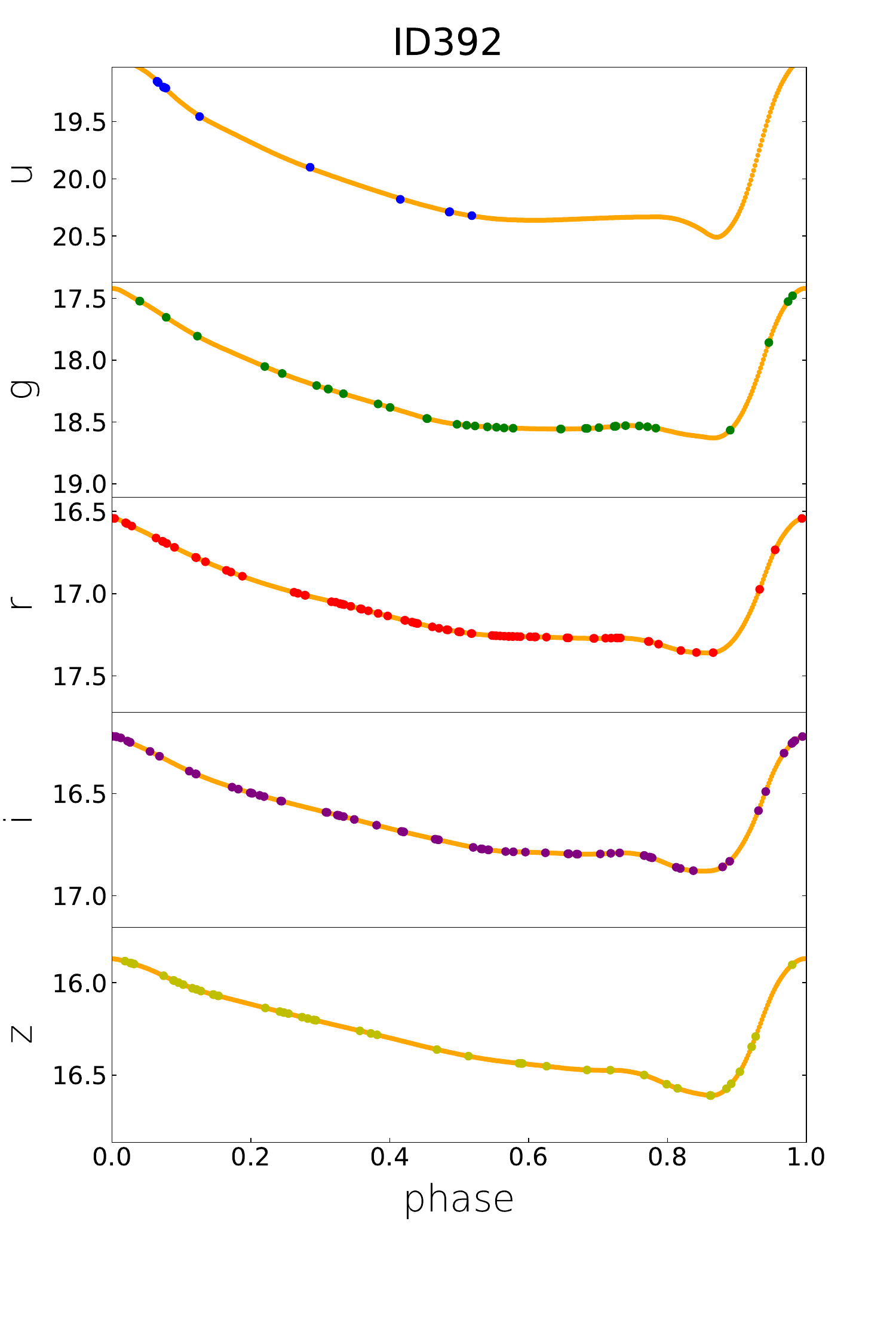}
\includegraphics[width=0.5\linewidth]{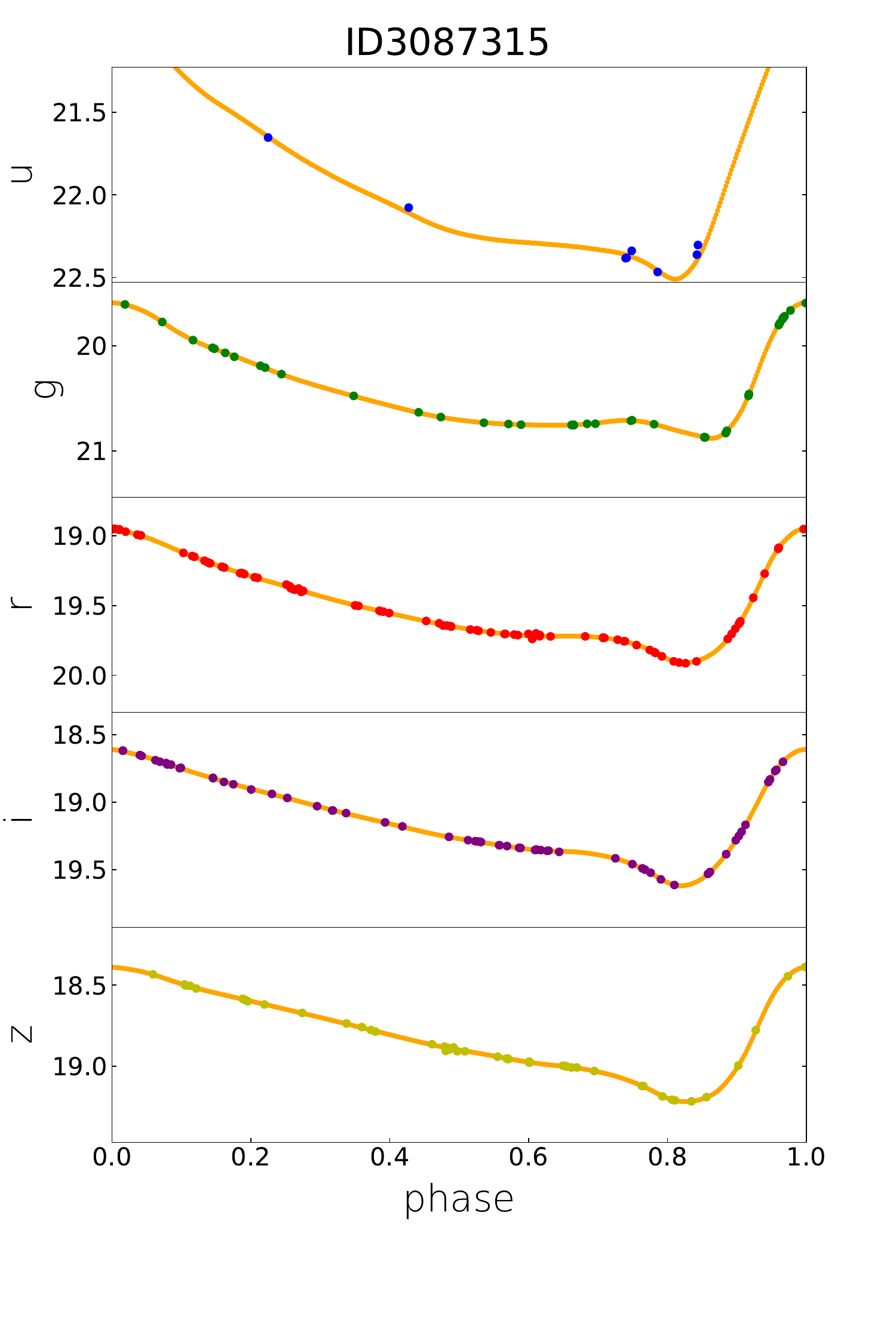}
\caption{Simulated LSST observations during the first 4 years in ugriz colors (colored points) for two RR Lyrae in Saha’s sample. Star 3087315 is the faintest in their sample. The orange light curve is Saha's template. Photometric errors (<0.001 mag) derived from Opsim and used in the thissimulation are too small to be shown.\label{figure1}}
\end{figure}
In Figure \ref{figure1}, we exhibit two out of the  simulated time series obtained in all available filters after 4 years of LSST observations, phase-aligned using the  periods derived by Saha (orange curve). It should be noted that unfortunately the Saha observations are missing the Y-band and for this reason, we did not simulate the observations in the Rubin-LSST y-band. The star ID3087315 is the faintest among the Saha's that, nevertheless, falls within Rubin detection capabilities ( single-visit   5-sigma magnitudes are  23.3, 24.4, 23.8, 23.6, and 22.9 in u,g,r,i,z filters respectively).
We have started selecting 4 years of the survey as the starting point for our investigation because according to the simulated LSST observational strategy, it is after this duration that, on average, the total number of visits will be comparable to those obtained by Saha ($\sim$ 250/300 phase points if all the  bands are considered and $\sim$ 60/70 in the 'i' band). Coincidentally, this is also the average number of visits expected for the low coverage of the   Galactic Plane regions (see Subsection 3.1).
\begin{figure}
\includegraphics[width=0.78\linewidth]{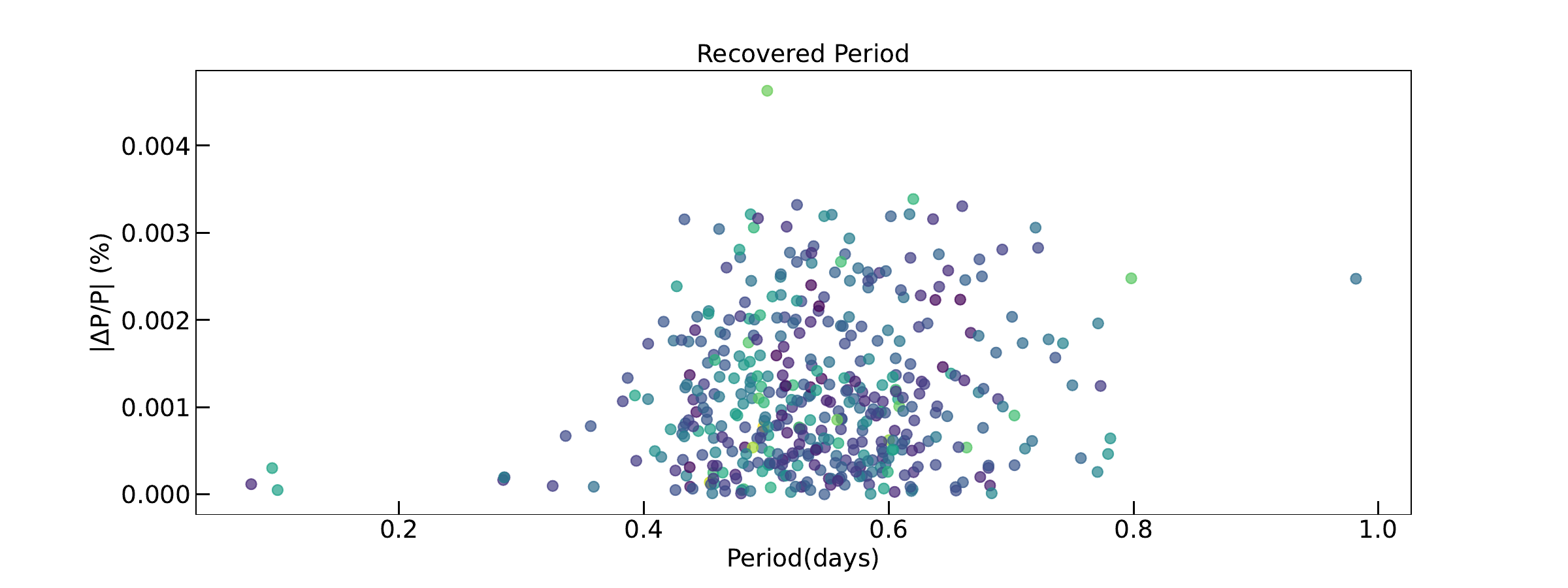}
\includegraphics[width=0.9\linewidth]{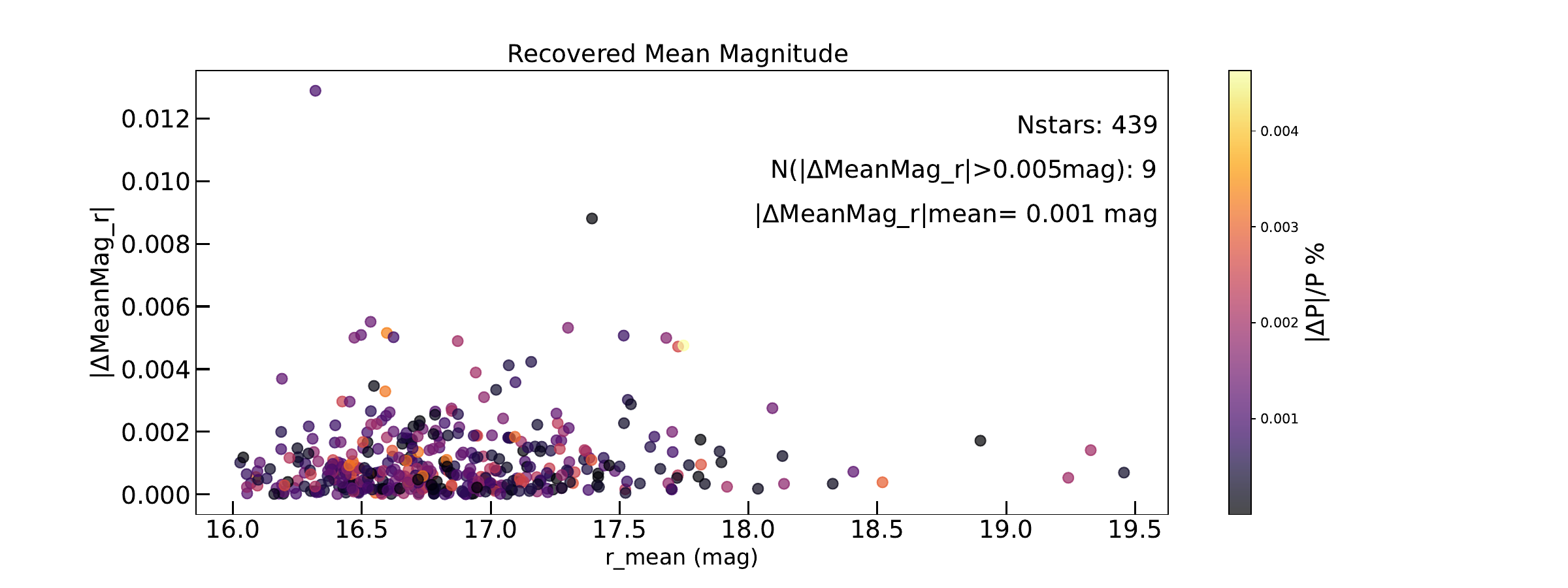}
\includegraphics[width=0.9\linewidth]{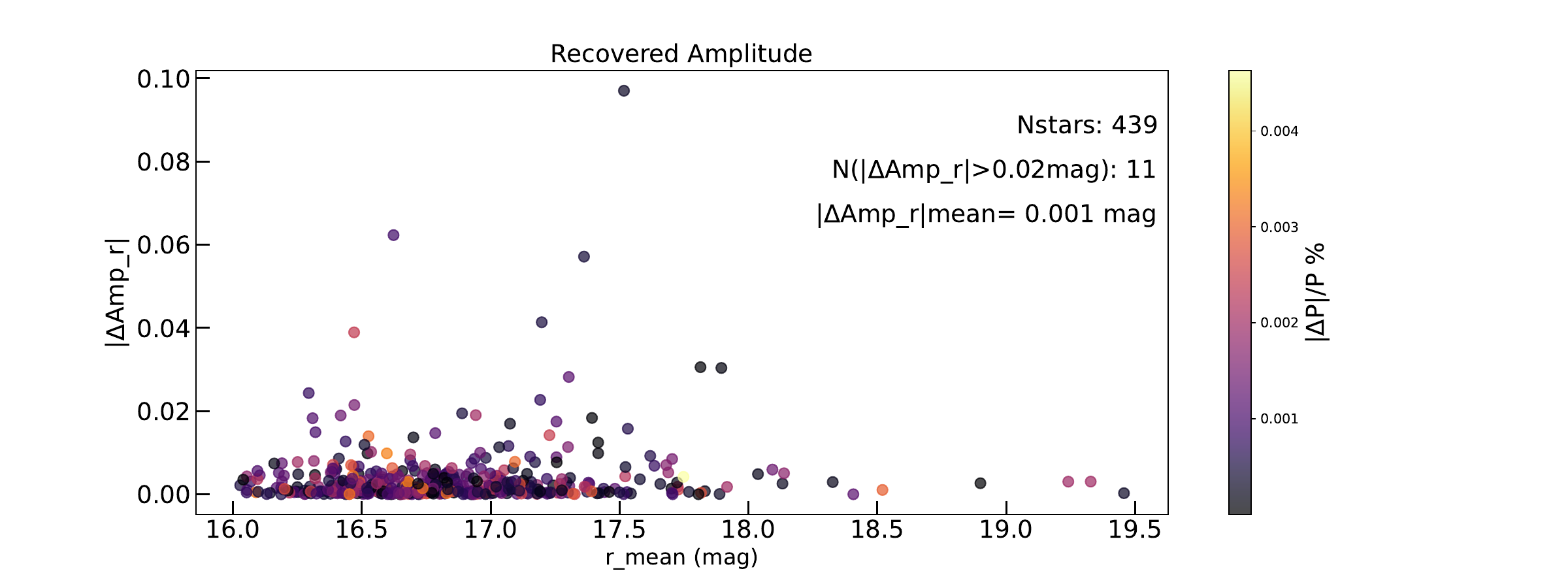}
\caption{Relative difference (in percentage) between recovered and template periods (upper), mean magnitude (middle), and amplitude (bottom) in function of r mean brightness for the 439 RRL from Saha's sample. 
In the middle and upper panels the vertical bar reports the accuracy of the recovery of the period. \label{figure3}}
\end{figure}
In the upper panel of Figure \ref{figure3}, we present the comparison between the periods derived by PSR metric on 4-year LSST simulated time series and Saha's periods.  We remember that our metric use Gatspy software  \citep{vanderplasivezic2015}  to derive periods. In particular we use the MultibandLombScargle option that  performs a Lomb-Scargle \citep{lomb, scargle} period search, by using simultaneously the information from all the available filters.
After 4 years, we found that the period is  recovered excellently: within 0.005$\%$  for all the stars and within 0.001$\%$ for more than the 50$\%$ of star's sample. 
The middle and bottom panels of Figure \ref{figure3},  display the trend with the template's mean magnitude of the difference between the recovered and template mean magnitudes (in r-band) after the phased light curve was fitted by our metric. The PSR metric's version used in this work  differs from the one published in \citet{dicriscienzo2023} as it autonomously computes the required numbers of harmonics, thus reducing the degrees of freedom in the fit.
Our analysis demonstrates that in the WFD survey after 4 years,  LSST  will recover Saha's light curves with such precision that it provides very accurate mean magnitudes and peak-to-peak amplitudes ($\sim$ 0.001 mag). It's crucial to note that these quantities, especially periods and mean magnitudes  go in the Period-Luminosity (PL) relations that are used to determine distance and/or metallicity. Thus, the high precision of the recovered parameters also implies extremely reliable distance and metallicity determinations. Insteed amplitude is less important for distances but it is relevant to study the properties of the RRL population through for example, a Bailey diagram \citep{fiorentino15, fiorentino22}.\\
The only error we have considered at this stage on the individual visit magnitude is the photometric error derived from the 5-sigma limiting magnitude given by Opsim at each position in the sky, which, notably, for the magnitudes involved, is exceedingly small (less than 0.006/0.007 mag in all the photometric bands). This factor also explains why our light curve exhibits much less dispersion around the template compared to Saha's light curve that exhibits a 0.03/0.04 mag mean scatter around the template in the griz photometric bands and   up to 0.09 mag in the u (see their Figure 1).\\ 
We didn't take into consideration the crowding effect on the template assuming that  \cite{saha19} accurately deblended their stars during their photometry. The effect of the crowding will be discussed in detail in Section 4.\\

\begin{figure}
\includegraphics[width=0.5\linewidth]{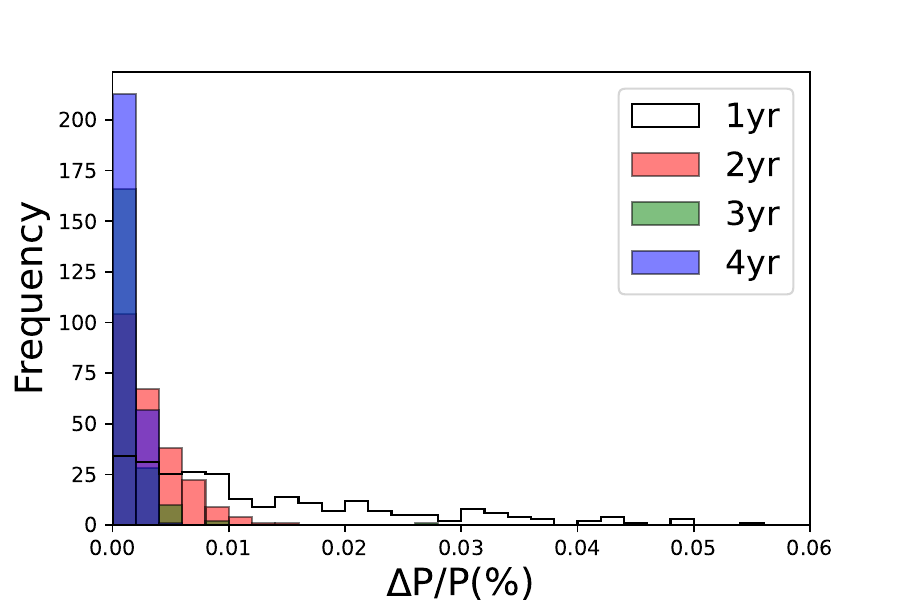}
\includegraphics[width=0.5\linewidth]{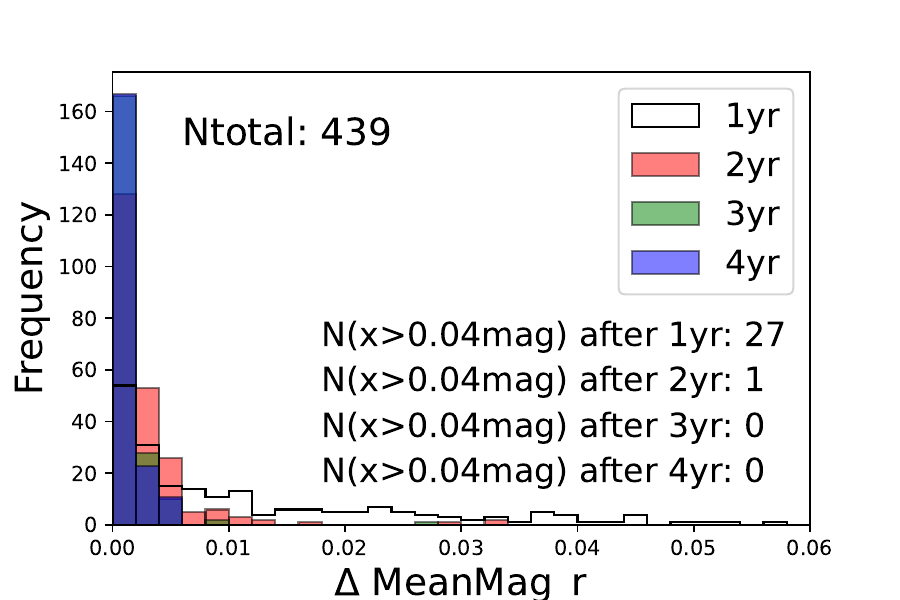}
\caption{Precision distribution in the recovery of period and average magnitude of RRL in the Saha sample after 1, 2, 3, and 4 years.The analysis of the second panel shows that starting from the second year, the precision in the r band remains below 0.04 mag.\label{figearly}}
\end{figure}
\subsection{Early Science: Unveiling Key Insights from the very first releases}
To investigate the LSST recovery of RRL light curves from the very first release, in Figure \ref{figearly}, we show the differences between recovered and true periods/mean magnitudes after 1, 2, 3, and 4 years. \\
Despite the first data release being expected six months after the survey's start, we didn't consider this release in the figure because in that case the number of visits strongly depends on the type of rolling conducted during the first part of the year a matter that, even though fundamental in defining LSST's early science and extensively discussed in the Phase2 Document,  falls beyond the scope of this work.\\ 
As expected,  our simulations show that the accuracy in recovering the light curve increases with the number of visit, but it is enough for doing early science even after one year. Entering in  more detail, we will recover periods with a precision smaller  0.005\% for almost half of the sample, while, in general, they will all remain below 0.05\% (that translates in about 4sec for a typical RRL ) from the first year. \\
Figure \ref{figearly} also shows that from the first year, r-average magnitudes will be recovered within 0.04 mag. Instead, the same precision in the g-band will be achieved by more than 90\% of the stars in the Saha sample only starting from the second year, while the number of visits in the u-band is too small (<7 phase points) to allow a good fit of the data. \\
Starting from the very first release, light curve templates (LCTs) can be adopted in this case to estimate accurate mean magnitudes (Braga et al. 2024, in prep.). LCTs can be used both by anchoring them on a single phase point (but this requires the knowledge of the reference epoch) or as fitting functions, for which at least three/four-phase points are needed, but only the pulsation period is required. This tool will be mostly useful during the first two years of the survey---when a precise fitting of the light curve will not be possible due to the small number of phase points---thus allowing  the exploitation of the early scientific data from LSST. 
In conclusion, with Rubin-LSST we will obtain light curves at least as accurate as those derived by \citet{saha19} from the early years of the Rubin-LSST survey, and additionally, over a wider and deeper footprint  with the large advantage of exploring a much larger portion of the Bulge.  For example, setting the threshold at 0.04 mag on the mean magnitude, which is the constraint imposed by \citet{bono} for the use of the REDIME method to simultaneously measure individual distances, reddening and metallicities of
RRL, our simulations show that these are achieved from the first year in the riz(and y), and from the second year in the g band. The visits in the u band within the first 4 years, however, are too few to allow for an adequate fit, and in that case, the use of templates will be necessary.\\
The analysis just described is also useful for understanding the problems related to the recovery of RRL's light curves located in the inner Disk region, which will be observed by LSST only in a low-density mode according to the last Opsim. Specifically, in 10 years of LSST, this region will receive only 250 visits, coincidentally the same number of visits after 4 years of WFD. 
If the filter balance remains the same as in the Wide Fast Deep survey, in the inner/bulge areas observed with a low density cadence we can expect that observations in the u-band for fitting the light curve will be practically useless even after 10 years. This makes it necessary to use templates in the u-band throughout the entire duration of the survey.

\section{Crowding effect on light curves }
So far, we have used Saha’s templates from \citet{seasar2010}  and analyzed their recovery without considering the effect caused by crowding.\\
Extending the analysis beyond the Baade Window and gooing deeper than the  Saha's variables,  which are the two main advantages of Rubin-LSST, also means having to tackle complicated issues such as crowding and reddening. In the upcoming sections, we will address these previously overlooked questions.\\
It is well known that photometry in a crowded field could lead to inaccurate measurement of the flux at each visit. In general, if the star is not well deblended from the neighboring stars, at each visit the measured flux will also include the flux from the surrounding stars. Naturally, this effect depends on both the density of stars in the field under consideration and the types (luminosity and color) of the blending stars. 
To show the impact of crowding on the recovery of light curves during LSST, we used a TRILEGAL simulation of the Galaxy described by \citet{daltio2022} that are made available at the NOIRLab Astro Data Lab \citep{olsen2018} .\\
\begin{figure}
\includegraphics[width=0.8\linewidth]{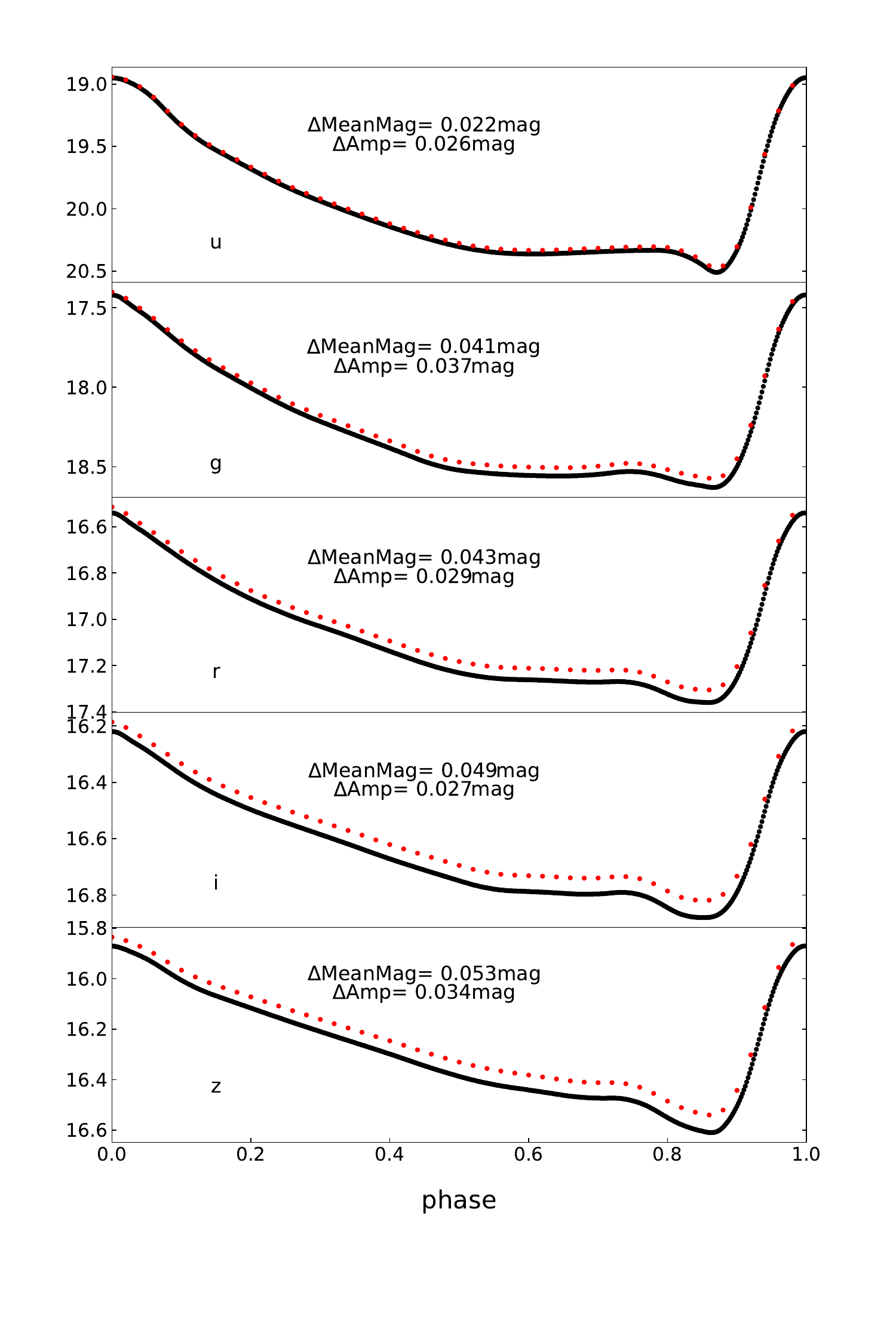}
\caption{\label{figure4} Comparison in ugriz bands between the Saha's light curve corresponding to the RRL star ID392 and the one obtained by adding the flux of 5 nearby stars according to TRILEGAL (see text). The difference between the average magnitudes and the amplitudes is reported in each panel.}

\end{figure}
In Figure \ref{figure4}, for example, we show the light curve of one of our sample's templates together with that obtained adding to each visit the flux of the stars falling within a radius equal to the LSST pixel scale.
In particular, for the  RRL shown in this figure (ID392) TRILEGAL finds 5 very faint stars with r-band magnitudes ranging from 21.35 mag and  26.68 mag.
In the worst-case scenario\footnote{Albeit unlikely because any photometry software would recognize these blending stars.},  where the brightness of nearby stars is detected as the brightness of the RRL  the effect would be to have light curves brighter by approximately 0.04 mag in r-band (see Figure \ref{figure4}). 
Another important effect is that the crowding decreases the RRL amplitude, due to its greater fractional contribution at minimum vs. maximum light. 

\begin{figure}
\includegraphics[width=0.7\linewidth]{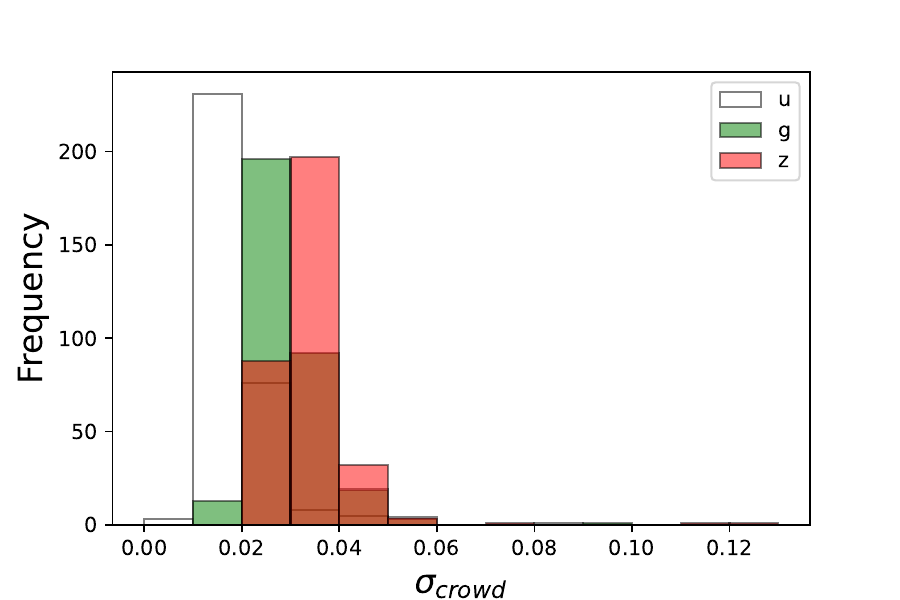}
\caption{
Histogram of photometric errors $\sigma_{crowd}$ due to stellar crowding derived using \texttt{CrowdingMagUncertMetric.py} for the RRL of the Saha sample in the labeled band.}
\label{figure5}
\end{figure}

By doing the same exercise for all the other stars in Saha's sample, which, being in different positions have different blending stars, the differences between recovered and template mean magnitudes remain for the investigated bands of the same order of magnitude, although they can reach up to one-tenth of a magnitude in cases where  blending stars are very luminous.\\
Obviously, providing a precise measure of the error due to crowding is a much more complex operation that requires  Artificial Star Tests (AST), where fake stars are added to the observed images and then recovered using the same pipeline used for the original images. 
Fortunately, approximate methods exist for determining the error function from simulations. One of this method is described by \citet{olsen2003} and  suggests that  the
photometric errors due to stellar crowding, $\sigma_{crowd}$, depend on an integral of the brightness distribution of stars fainter than the faintest one being observed. This method is actually used in the MAF metric \texttt{CrowdingMagUncertMetric} developed by Girardi and collaborators and described in \citet{daltio2022}.
In Figure \ref{figure5}, we present the values obtained with this metric taking into account positions and magnitudes (in the labeled bands) for the objects in the Saha's sample. It's noteworthy that not only that these errors are of the same order of magnitude as the scatter observed in Saha's light curves, but also that they are larger than the precision on the average magnitudes we obtained from the recovery of the simulated light curves.
These  computed crowding errors should be considered as errors on the zero point of the recovered light curve (in the brighter direction). Therefore these errors should be associated  to the mean magnitude obtained from the light curve fitting together with \texttt{$\Delta$MeanMag} derived  by PSR metric.\\
LSST Data Release is not planning to include
estimates of incompleteness and photometric errors due to crowding (see LSST Data Products Definition Document (  \url{https://lse-163.lsst.io/}). 
As demonstrated in this section these estimates are essential  to understand the uncertainty obtained in the recovery of light curves of pulsating stars, but more generally these estimates are necessary to
transform star counts into absolute quantities such as stellar volume densities and lifetimes in different phases of the stellar evolution, and to be able to fit the physical models to the stellar data.
For this reason, we strongly suggest that crowding errors are a solid added value to be  associated with LSST Data Releases.

\section{Probing the far  side of the Bulge with LSST }
\begin{figure}
\includegraphics[width=0.9\linewidth]{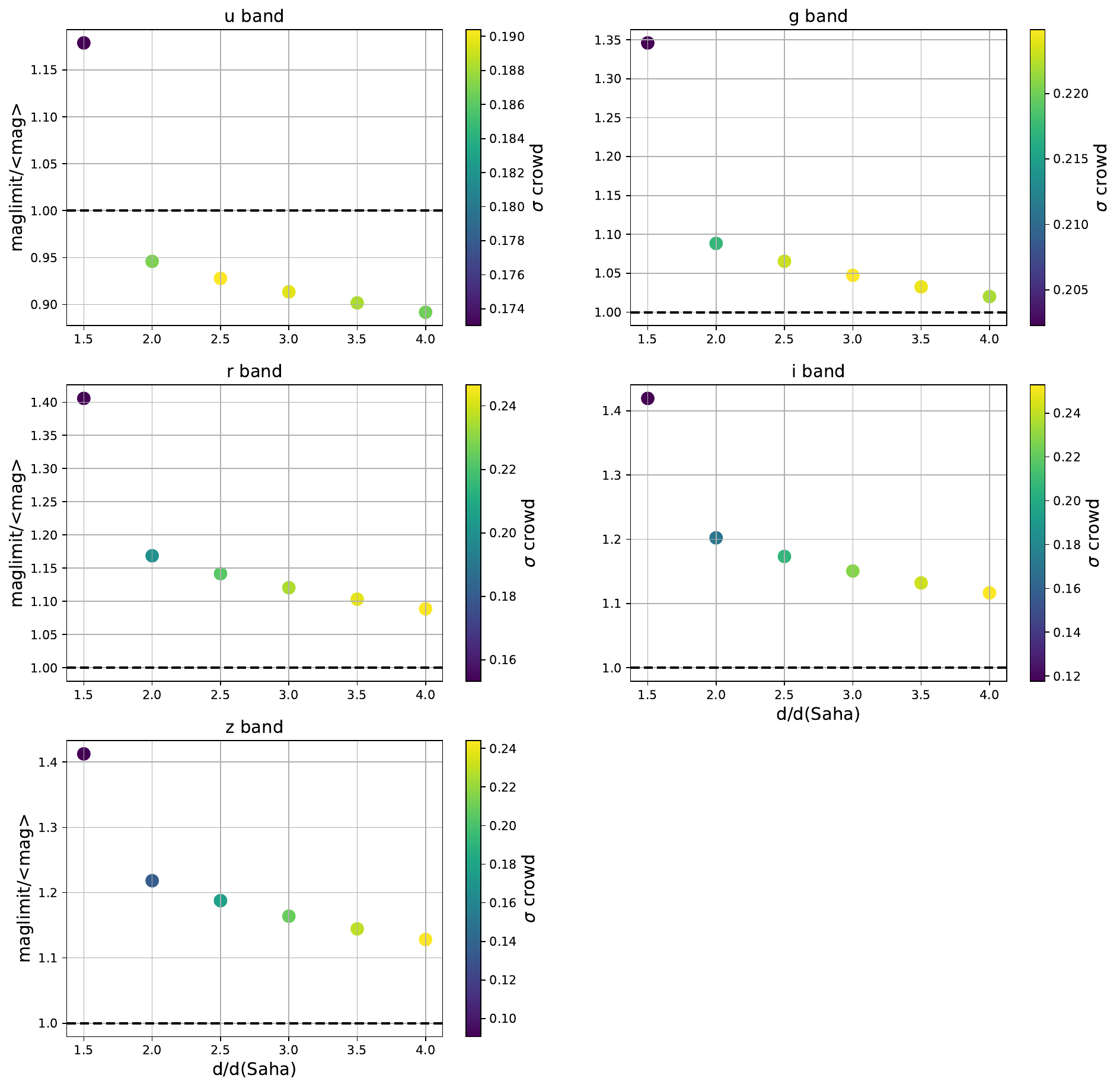}
\caption{\label{putfurther}Distance of the template's average magnitude from the limiting magnitude derived from Opsim for different template distances expressed as factors between new distance and original distance derived from Saha color-coded by the $\sigma_{crowd}$ value.  The points below the horizontal line are unlikely to be observed by LSST according to the most recent simulations.}
\end{figure}
So far, we have explored how Rubin-LSST will be able to recover Saha’s LC RRL over time and the level of precision that will be achieved with each release, aware that the advantage of Rubin-LSST will be that of extending Saha’s work to a much larger portion of the sky than that observed by DECam.
In this section instead, we investigate to which photometric depth can RRL be observed/recovered.
This is a crucial information for understanding our ability to reconstruct the geometry and formation history of the bulge using LSST data.\\
To this purpose, 
we take   star ID392 that according Saha's computations is located  at $d_{saha}$=9.09Kpc and has a color excess E(r-z)=0.86 mag, and while staying within the Baade Window, we move the star away using different ratios of  $d_{saha}$ which thus becomes  our distance unit.
To account for the increase in reddening  an additional color excess had to be added to each new distance to derive the simulated light curve. This additional color excess takes into account the high probability to encounter a significant amount of dust in the other side of the bulge 
Specifically, for all the distances larger than $d_{saha}$  we assumed that the additional excess is equal to the quantity calculated by Saha using the minimum light curve method \citep{sturch} on our side. This assumption is based on the spherical symmetry of the dust around the Galactic center, with a drop to zero when radius R=$d_{saha}$. We acknowledge that this is only an approximation but without access to models of dust distribution beyond the bulge\footnote{For example, we verified that the 3D dust Bayestar-maps  \citep{green2018} saturate at distances greater than the Galactic center in the direction of ID392}, it is the only plausible approach. \\
Actually, the awareness of  this uncertainty makes still more evident  the importance of multi-wavelength LSST observations of RR Lyrae stars towards the Bulge, will help to derive reddening maps, either using the appropriate combination of theoretical relations from \citet{marconi2022} as suggested by the REDIME method \citep{bono}, or directly from RRL LCs, as done by \citet{saha19}, using the color at minimum light of the RRL.
Figure \ref{putfurther} separately shows, for each band, the ratio between the  limiting magnitude (derived from Opsim) and the ID392's mean magnitude  as the distance of the star increases. Obviously, due to the increased reddening in the u-band\footnote{In particular we adopt the A$_X/$E(r-z) ratios as given in Equations (22)–(26) of \citet{saha19}} this limit in the bluest available photometric band is reached at shorter distances with respect of the others. In particular, for the considered extra reddening in the specific case, doubling the distance already brings the magnitudes below the photometric threshold. For the other bands, however, the figure shows that in principle it is possible to observe stars up to four times the original distance, roughly up to just under $\sim$40 kpc. However, at these distances, the so-called \textit{crowding limit} (the brightness below which the incompleteness caused by crowding becomes significant, assuming values  above approximately 50$\%$) is reached. As reported by \citet{daltio2022} this closely corresponds to the point where the near-infrared magnitude will first reach $\sigma_{crowd}$ $\sim$ 0.25 mag. This limit also depends on the instantaneous value of the seeing, in particular, for the case in question,  we used the minimum seeing in the first four years of simulations (0.51") and for this reason, the values plotted in the figure represent a lower limit. This is true especially for the near-infrared bands riz (and y when it will be available), while in the case of the g-band, this limit is reached before, and in particular at three times the distance of Saha.\\
For all distances between the true distance of the template and the limit distance due to the crowding, LSST provides valuable data for discovering new RR Lyrae stars. The time series of these newly discovered variables will allow us to achieve excellent accuracy on the period (up to $\Delta$P=10$^{-6}$days). We are also able to reconstruct the shape of the light curve and hence to derive its zero-point representing the mean magnitude. But for the errors on this value, as stressed in the previous section, we cannot consider only those coming from the fitting procedure, but we have to take into account the larger uncertainties due to the effect of crowding on the zero-point of the light curve. In our test case, for example, $\sigma_{crowd}$ ranges from 0.2 mag in the g-band to 0.1 mag in the z-band aiming to make the mean magnitudes brighter. 
If these errors are not properly taken into account we would derive, on average, smaller distances of those RRL observed in the opposite side of the Bulge that are brighter than the crowding limit, leading to an incorrect reconstruction of the  shape of the Bulge.
Unfortunately, even at 1.5 times the distance of Saha ($\sim$13.5 kpc), the errors are still significant even if they drop below 0.1mag in our reddest band z, and presumably remain equally small even in y band, which we are currently not simulating because we lack the corresponding  Saha's template. 
Therefore, we confirm that these two bands will be crucial for the first kpc beyond the Bulge to characterize their LC  and derive the RRL pulsation parameters such as periods and  mean magnitudes. Unfortunately, having mean magnitudes in only two bands means that it won't be possible to use the REDIMI method to simultaneously derive distance, metallicity, and reddening. Different methods based on the shape of the light curve, such as $\phi$31 –[Fe/H] relations, could be used in these cases (Ripepi et al. in preparation)

\section{Conclusions}

In \citet{dicriscienzo2023}, we employed the tool \texttt{PulsationalStarRecovety} developed by our team during the Survey Strategy Optimization process to simulate Rubin- LSST observations of RRL in selected Local Group dwarf galaxies and quantified the recovery of their light curves. We found that the observing cadence will not have a major impact on the detection and characterization of individual RRL, however, it will affect the number of RRL measured with the required photometric accuracy for properly utilizing \citet{marconi2022} relations to derive their distances and metallicities from observational quantities such as period and mean magnitudes.\\
In this paper, our focus shifted towards a significantly more complex footprint: the Bulge, which was only recently included in  Rubin-LSST main survey and for which there are still many uncertainties regarding both the choice of the exact size of the footprint  and the observational strategy to be used.\\
In this work, we have shown that high reddening and crowding of this part of the sky will affect photometric measurements during each survey visit and pose challenges to the recovery of pulsating stars' light curves, significantly increasing the error budget on pulsation parameters especially on  average magnitudes and  amplitudes.\\
The same considerations can obviously be extended to observations of other crowded fields such as  globular clusters or the central parts of dwarf galaxies studied in \citet{dicriscienzo2023}.\\
As a reference point for this work, we relied on Saha's observations with CTIO/DECam towards the Bulge. In \citet{saha19} the authors used these observations to reconstruct the spatial distribution of  RRL population around the Galactic center and at the same time derived detailed reddening maps. Furthermore, these observations can be combined with \citet{marconi2022}  relations to determine the metallicity distribution of observed RRL using the  REDIME method by \citet{bono} (Braga et al. in preparation) or reddening maps based  on the color at minimum light of RRL.
Currently  this is  possible only in a limited area towards the Bulge observed by DECam, whereas Rubin-LSST will give the opportunity  to perform  similar analysis over a much broader and deeper area providing homogeneous material for future studies of the Bulge structure, the first step in reconstructing its formation history.\\
In particular, our simulations show that although it will be necessary to use variable templates to accurately derive observables such as mean magnitude and amplitude, especially in undersampled photometric bands like u and g during the very early releases, using the period derived from the most sampled bands, on the contrary for the redder  bands starting from DR1  the precision achieved in the LC recovery will be much higher than that obtained by Saha and collaborators. Instead, in the case of Bulge/Disk  regions covered by the WFD survey we will have to wait for the data release at the end of the second year to achieve the same precision even in the u and g bands without using templates and even the end of the survey in the case of areas not included in the main survey.\\
Using  TRILEGAL’s simulated Galaxy by \citet{daltio2022} we have analyzed the effect of crowding on the recovery of RRL light curves  emphasizing  that if not appropriately treated, we recover  light curves that are brighter and with smaller amplitude.
In particular  the need for a measure of crowding error starting from the early releases has been emphasized, in order to properly utilize the provided magnitudes appropriately.\\
We finally explored the possibilities of observing RR Lyrae stars in  the opposite side of the Bulge with respect to the Galactic center. Obviously, the results depend heavily on the reddening model adopted for the dark side of the Bulge. Specifically, by making the very approximate assumption of a uniform radial distribution that  decreases to zero after a radius R=$d_{saha}$, we have demonstrated that: 1) observations in the u  band fall below the photometric limit even before reaching R=2$d_{saha}$, therefore before reaching virtually the point symmetrically opposite to us, 2) in the infrared photometric bands, the crowding limit is reached before the photometric limit, and this occurs at around 4 times the distance of the test star we used. In this case, the recovery of light curves, although it will provide very precise periods and equally good light curve shapes, will give average magnitudes that will be heavily influenced by the crowding limits (between 0.1mag and 0.2 mag). \\
Among the investigated bands the only one to obtain distance estimates accurate enough to investigate the shape and fine structure of the Bulge/Bar regions well beyond the galactic center is the z-band. Additionally, we note that the y-band will play a crucial role in early science, especially in crowded and reddened regions like the inner bulge. However, constraining the photometric accuracy in this band requires further refinement. \\
In general, early Rubin-LSST observations will be able to trace the edge between the Bulge/Bar and the thin Disk, which is a very promising indication. Additionally, these observations will help in tracing the transition between the Bulge/Bar and the thin disk. In a recent investigation by our group \citep{dorazi24}, a significant sample of field RRL with disk kinematics was identified. Despite their similarity to typical thin Disk stars, these objects exhibit systematic under abundance in $\alpha$ and neutron capture elements compared to typical thin Disk stars. Once confirmed, if these findings hold true, they could pave the way for a new approach to constrain the early formation of the Galactic spheroid, particularly the early coupling between the Bulge/Bar and the Thin disk, using the same stellar tracer.
In conclusion, we strongly support the inclusion of Bulge/Inner Disk observations in the WFD because LSST observations of this part of the sky, despite the limitations discussed in this paper, will be crucial for reconstructing the oldest component of the Bulge, thanks to the RR Lyrae stars that will be discovered and analyzed during the survey.

\begin{acknowledgments}
This work was supported by: “Preparing for Astrophysics with the LSST Program" funded by the Heising-Simons Foundation, and administered by Las Cumbres Observatory with a grant for the publication and with the Kickstarter grant “Period and shape recovery of light curves of pulsating stars in different Galactic environments (KSI-8)”;  Mini grant INAF 2022 “MOVIE@Rubin-LSST: enabling early science" (PI: Di Criscienzo, M.); Project PRIN MUR 2022 (code 2022ARWP9C) “Early Formation and Evolution of Bulge and HalO (EFEBHO)" (PI: Marconi, M.),  funded by European Union – Next Generation EU; Large grant INAF 2023 MOVIE (PI: M. Marconi). VB and RC thank the Rubin-LSST inkind ITA-INAF-S22  for the support.
M.M. acknowledges support from Spanish Ministry of Science, Innovation and Universities (MICIU) through the Spanish State Research Agency under the grants "RR Lyrae stars, a lighthouse to distant galaxies and early galaxy evolution" and the European Regional Development Fun (ERDF) with reference PID2021-127042OB-I00 and from the Severo Ochoa Programe 2020-2023 (CEX2019-000920-S).
\end{acknowledgments}

\end{document}